\documentclass{elsart}
\usepackage{graphicx}
\usepackage{amssymb}
\begin{document}
\begin{frontmatter}
\title{How Required Reserve Ratio Affects Distribution and Velocity of Money}
\thanks{The primary version of this paper has been
presented at the Third Nikkei Econophysics Research Workshop and
Symposium.}
\author{Ning Xi, Ning Ding, Yougui Wang\corauthref{cor1}}

\corauth[cor1]{Corresponding author.
\\ {Tel.: +86-10-58807876; fax:+86-10-58807876.}
\\{\em E-mail address:\/}\, ygwang@bnu.edu.cn (Y. Wang)}
\address{Department of Systems Science, School of Management, Beijing Normal University, Beijing,
100875, People's Republic of China}

\begin{abstract}
In this paper the dependence of wealth distribution and the
velocity of money on the required reserve ratio is examined based
on a random transfer model of money and computer simulations. A
fractional reserve banking system is introduced to the model where
money creation can be achieved by bank loans and the monetary
aggregate is determined by the monetary base and the required
reserve ratio. It is shown that monetary wealth follows asymmetric
Laplace distribution and latency time of money follows exponential
distribution. The expression of monetary wealth distribution and
that of the velocity of money in terms of the required reserve
ratio are presented in a good agreement with simulation results.
\end{abstract}


\end{frontmatter}
\section{Introduction}

Recently two essential aspects of money circulation have been
investigated based on the random transfer models
\cite{redner,Bouchaud,basic,saving1,saving2,hayes,circu,prefer}.
One is statistical distribution of money, which is closely related
to earlier Pareto income distribution \cite{pareto} and some
recent empirical observations \cite{emp1,emp2,emp3,emp4,emp5}. The
other one is the velocity of money, which measures the ratio of
transaction volume to the money stock in an economic system. All
the models which appeared in these researches regarded the
monetary system as being composed of agents and money, and money
could be transferred randomly among agents. In such a random
transferring process, money is always being held by agents, any
single agent's amount of money may strongly fluctuate over time,
but the overall equilibrium probability distribution can be
observed under some conditions. The shape of money distribution in
each model is determined by its transferring rule, for instance,
random exchange can lead to a Boltzmann-Gibbs distribution
\cite{basic}, transferring with uniform saving factor can lead to
a Gaussian-like distribution \cite{saving1} and that with diverse
saving factors leads to a Pareto distribution \cite{saving2}. On
the other hand, the time interval between two transfers named as
holding time of money is also a random variable with a steady
probability distribution in the random transferring process. The
velocity of money could be expressed as the expectation of the
reciprocal of holding time and the probability distribution over
holding time was found to follow exponential or power laws
\cite{circu,prefer}.

The amount of money held by agents was limited to be non-negative
in the models mentioned above except Ref. \cite{basic}. Allowing
agents to go into debt and putting a limit on the maximal debt of
an agent, Adrian Dr\u{a}gulescu and Victor Yakovenko demonstrated
the equilibrium probability distribution of money still follows
the Boltzmann-Gibbs law. Although they devote only one section to
discussing the role of debt in the formation of the distribution,
they are undoubtedly pathfinders on this aspect. As cited in their
paper, ``debts create money" \cite{econo1}. Specifically, most
part of the money stock is created by debts through banking
system, and this process of money creation plays a significant
role in performance of economy especially by affecting the
aggregate output \cite{econo2}. Thus money creation should not be
excluded from discussion on the issues of monetary economic
system. With cognition of this significance, Robert Fischer and
Dieter Braun analyzed the process of creation and annihilation of
money from a mechanical perspective by proposing analogies between
assets and the positive momentum of particles and between
liabilities and the negative one \cite{Braun1,Braun2}. They
further applied this approach into the study on statistical
mechanics of money \cite{Braun3}.

As well known, the central bank plays an important role of
controlling the monetary aggregate that circulates in the modern
economy. It issues the monetary base which is much less than the
monetary aggregate. The ratio of the monetary aggregate to the
monetary base is called the money multiplier. The central bank
controls the monetary aggregate mainly by adjusting the monetary
base and by setting the required reserve ratio which is a key
determinant of the multiplier. So the required reserve ratio is
crucial in monetary economic system. The aim of this work is to
investigate the impacts of the required reserve ratio on monetary
wealth distribution and the velocity of money. Our model is an
extended version of that of Robert Fischer and Dieter Braun
\cite{Braun3}. In their model, random transfer would increase the
quantity of money without bounds unless some limits are imposed
exogenously on the stock of assets and liabilities, which are
given by specifying an aggregate limit or imposing a transfer
potential. Compared with this, we introduce the monetary base and
the required reserve ratio in our model by interpreting the
process of money creation with the simplified money multiplier
model. Thus the limit can be governed by setting the initial
values of the monetary base and the required reserve ratio. In
addition, we adopt the conventional economic definition of money
instead of what they used. We think that the conventional
definition of money is more appropriate to the analysis of
realistic monetary system. We hope that our work can expose the
role of the required reserve ratio in monetary circulation and is
helpful to understand the effect of the central bank on monetary
economic system.

This paper is organized as follows. In next section we make a
brief presentation of money creation and the simplified multiplier
model. In Section $3$ we propose a random transfer model of money
with a bank. And the shapes of monetary wealth distribution and
latency time distribution are demonstrated. In Section $4$ the
dependence of monetary wealth distribution and the velocity of
money on the required reserve ratio is presented quantitatively.
We finish with some conclusions in Section $5$.

\section{Money Creation and Simplified Multiplier Model}
Modern banking system is a fractional reserve banking system,
which absorbs savers' deposits and loans to borrowers. Generally
the public holds both currency and deposits. As purchasing, the
public can pay in currency or in deposits. In this sense, currency
held by the public and deposits in bank can both play the role of
exchange medium. Thus the monetary aggregate is measured by the
sum of currency held by the public and deposits in bank in
economics. When the public saves a part of their currency into
commercial banks, this part of currency turns into deposits and
the monetary aggregate does not change. Once commercial banks loan
to borrowers, usually in deposit form, deposits in bank increase
and currency held by the public keeps constant. So loaning
behavior of commercial banks increases the monetary aggregate and
achieves money creation.

Money creation of commercial banks is partly determined by the
required reserve ratio. In reality, commercial banks always hold
some currency as reserves in order to repay savers on demand.
Total reserves are made up of ones that the central bank compels
commercial banks to hold, called required reserves, and extra ones
that commercial banks elect to hold, called excess reserves.
Instead of appointing required reserves for each of commercial
banks, the central bank specifies a percentage of deposits that
commercial banks must hold as reserves, which is known as the
required reserve ratio. The role of the required reserve ratio in
money creation is illuminated well by the multiplier model
\cite{multi0}.

The multiplier model, originally developed by Brunner and Meltzer
\cite{multi1,multi2}, has become the standard paradigm in the
textbooks of macroeconomics. We introduce its simplified version
here. In monetary economic system, the monetary aggregate can be
measured by
\begin{equation}
M=C+D \label{aggregate},
\end{equation}
where $C$ denotes currency held by the public and $D$ denotes
total deposits. The monetary base $M_{0}$ is the sum of currency
held by the public and reserves in the banking system, $R$:
\begin{equation}
M_{0}=C+R \label{base}.
\end{equation}%
Reserves, which are decomposed into required reserves $RR$ and
excess reserves $ER$, can be given by
\begin{equation}
R=RR+ER \label{reserve1}.
\end{equation}%
Required reserves can be calculated according to the required
reserve ratio $r$ and deposits in commercial banks $D$:
\begin{equation}
RR=rD \label{RR}.
\end{equation}%
So Equation (\ref{reserve1}) can be rewritten as
\begin{equation}
R=ER+rD \label{reserve2}.
\end{equation}%
For simplicity, assume that the public holds no currency in hand
and that excess reserves are always zero. With these two
assumptions, combining Equations (\ref{aggregate}), (\ref{base})
and (\ref{reserve2}) produces the monetary base-multiplier
representation of the monetary aggregate:
\begin{equation}
M=mM_{0} \label{MA},
\end{equation}%
where $m$, the money multiplier, is given by
\begin{equation}
m=\frac{1}{r} \label{multi}.
\end{equation}%
According to this representation, an increment of one dollar in
the monetary base produces an increment of $1/r$ dollars in the
monetary aggregate. Since loans made by commercial banks create
equal amount of money, its volume $L$ is the difference between
the monetary aggregate and the monetary base, that is
\begin{equation}
L=\frac{M_{0}}{r}-M_{0} \label{loan}.
\end{equation}%
This equation shows clearly the relation between money creation
and the required reserve ratio. As the required reserve ratio
increases, the capability of money creation declines. Please note
if the public holds currency in hand or commercial banks decide to
keep some amount of currency as excess reserves, the amount of
money $L$ created by the banking system is less than the value
given by the right-hand side of Equation (\ref{loan}).

Although all factors involved in money creation except the
required reserve ratio are ignored in the simplified multiplier
model, it conveys us the essence of money creation in reality.
This suggests that the role of money creation can be investigated
by focusing on the impacts of the required reserve ratio on
relevant issues. Thus we simply introduced a bank into the random
transfer model to examine how the required reserve ratio affects
monetary wealth distribution and the velocity of money.

\section{Model and Simulation}

Our model is an extension of the model in Ref. \cite{Braun3}. The
economy turns into the one consisting of $N$ traders and a virtual
bank. We postulate that all traders hold money only in deposit
form throughout the simulations. At the beginning, a constant
monetary base $M_{0}$ is equally allocated to $N$ traders and is
all saved in the bank. As a result, total reserves held by the
bank are $M_{0}$ at the beginning. Time is discrete. Each of the
traders chooses his partner randomly in each round, and yield $N$
trade pairs. In each trade pair, one is chosen as ``payer''
randomly and the other as ``receiver''. If the payer has deposits
in the bank, he pays one unit of money to the receiver in deposit
form. If the payer has no deposit and the bank has excess
reserves, the payer borrows one unit of money from the bank and
pays it to the receiver in deposit form. But if the bank has no
excess reserve, the trade is cancelled. After receiving one unit
of money, if the receiver has loans, he repays his loans.
Otherwise the receiver holds this unit of money in deposit form.

Simulations are expected to show the results of two issues. One is
monetary wealth distributions. Monetary wealth is defined as the
difference between deposit volume and loan volume of a trader.
Thus the data of deposit and loan volumes of each trader need to
be collected. The other is the velocity of money. When the
transferring process of currency is a Poisson process, the
velocity of money can be calculated by latency time, which is
defined as the time interval between the sampling moment and the
moment when money takes part in trade after the sampling moment
for the first time. The distribution of latency time in this case
takes the following form
\begin{equation}\label{latency time}
    P(t)=\frac{1}{T}e^{-\frac{t}{T}} \label{latency},
\end{equation}
where $1/T$ is the intensity of the Poisson process. It can be
obtained by simple manipulation that the velocity of money is the
same as the intensity \cite{circu}. Thus we have,
\begin{equation}\label{velocity1}
    V=\frac{1}{T}.
\end{equation}
As collecting latency time, each transfer of the deposits can be
regarded as that of currency chosen randomly from reserves in the
bank equivalently.

Since the initial settings of the amount of money and the number
of traders have no impacts on the final results, we performed
several simulations with $M_{0}=2.5\times10^{5}$ and
$N=2.5\times10^{4}$, while altering the required reserve ratio. It
is found that given a required reserve ratio the monetary
aggregate increases approximately linearly for a period, and after
that it approaches and remains at a steady value, as shown in
Figure $1$.  We first recorded the steady values of the monetary
aggregate for different required reserve ratios and the results
are shown in Figure $2$. This relation is in a good agreement with
that drawn from the simplified multiplier model. We also plotted
the values of time when the monetary aggregate begins to be steady
for different required reserve ratios in Figure $3$. Since the
maximal value among them is $1.2\times10^{5}$ or so, the data of
deposit volume, loan volume and latency time were collected after
$8\times10^{5}$ rounds. We are fully convinced that the whole
economic system has reached a stationary state by that moment.

As shown in Figure $4$, monetary wealth is found to follow
asymmetric Laplace distribution which is divided into two
exponential parts by Y Axis \cite{asymmetric}, which can be
expressed as $p_{-}(m){\propto}e^{\frac{m}{\overline{m}_{-}}}$ and
$p_{+}(m){\propto}e^{-\frac{m}{\overline{m}_{+}}}$ respectively,
where $\overline{m}_{+}$ is the average amount of positive
monetary wealth and $\overline{m}_{-}$ is the average amount of
negative monetary wealth. This asymmetry of the distribution
arises from the non-zero monetary base set initially in our model
which money creation can be achieved on the basis of. It is worth
mentioning that in Ref. \cite{Braun3} the distribution with such a
shape can also be obtained by imposing an asymmetric,
triangular-shaped transfer potential. From simulation results, it
is also seen that latency time follows an exponential law, as
shown in Figure $5$. This result indicates that the transferring
process of currency is indeed a Poisson type.

\section{Results and Discussion}

\subsection{Monetary Wealth Distribution Versus the Required Reserve Ratio}\label{money dis}

We show monetary wealth distributions for different required
reserve ratios in Figure $6$. It is seen that both
$\overline{m}_{+}$ and $\overline{m}_{-}$ decrease as the required
reserve ratio increases. When the required reserve ratio increases
closely to $1$, $\overline{m}_{-}$ decreases closely to $0$ and
the distribution changes gradually from asymmetric Laplace
distribution to Boltzmann-Gibbs law which is the result from the
model of Adrian Dr\u{a}gulescu and Victor Yakovenko.

The stationary distribution of monetary wealth can be obtained by
the method of the most probable distribution \cite{probable}. In
our model, if $N$ traders are distributed over monetary wealth,
with $n_{m}$ traders holding monetary wealth $m (\geq 0)$,
$n_{m'}$ traders holding monetary wealth $m' (<0)$, this
distribution can be done in
\begin{equation}\label{microstate}
    {\Omega}=\displaystyle\frac{N!}{\displaystyle{\displaystyle\prod\limits_{m}}n_{m}!{\prod\limits_{m'}}n_{m'}!}
\end{equation}
ways. It is also required that the total number of traders, the
total amount of positive monetary wealth $M_{+}$ and that of
negative monetary wealth $M_{-}$ must be kept constant at
stationary state, that is
\begin{equation}\label{eq:trader}
    N={\sum_{m}}n_{m}+{\sum_{m'}}n_{m'},
\end{equation}
\begin{equation}\label{eq:positive}
    M_{+}={\sum_{m}}n_{m}m=\frac{M_{0}}{r}
\end{equation}
and
\begin{equation}\label{eq:negative}
    M_{-}={\sum_{m'}}n_{m'}m'=M_{0}-\frac{M_{0}}{r}.
\end{equation}
The stationary distribution can be obtained by maximizing
$\ln\Omega$ subject to the constraints listed above. Using the
method of Lagrange multipliers, we have
\begin{equation}\label{eq:lagrange}
    \d\ln\Omega-\alpha\d N-\beta\d M_{+}-\gamma\d M_{-}=0,
\end{equation}
whose solutions can be given respectively by
\begin{equation}\label{eq:soluion+}
    n_{m}=e^{-\alpha-{\beta}m}
\end{equation}
and
\begin{equation}\label{eq:solution-}
    n_{m'}=e^{-\alpha-{\gamma}m'}.
\end{equation}
So the stationary distribution can be expressed in continuous form
as
\begin{equation}\label{eq:distribution}
\begin{array}{ll}
    p_{+}(m)=\displaystyle\frac{N_{0}}{N}e^{-\beta m} \quad & \textrm{for $m \geq 0$;} \\
    \\[-18pt]
    p_{-}(m)=\displaystyle\frac{N_{0}}{N}e^{-\gamma m} \quad & \textrm{for $m < 0$,} \\
\end{array}
\end{equation}

where $N_{0}=e^{-\alpha}$ denotes the the number of traders with
no monetary wealth. Substituting Equations (\ref{eq:soluion+}) and
(\ref{eq:solution-}) into Equations (\ref{eq:trader}),
(\ref{eq:positive}) and (\ref{eq:negative}), and replacing
summation symbol with integral one, we have
\begin{equation}\label{eq:con1}
    (\frac{1}{\beta}-\frac{1}{\gamma})e^{-\alpha}=N,
\end{equation}
\begin{equation}\label{eq:con2}
    \frac{1}{{\beta}^2}e^{-\alpha}=\frac{M_{0}}{r}
\end{equation}
and
\begin{equation}\label{eq:con3}
    -\frac{1}{{\gamma}^2}e^{-\alpha}=M_{0}-\frac{M_{0}}{r},
\end{equation}
where Equation (\ref{eq:con1}) holds only when $\beta>0$ and
$\gamma<0$. Combining Equations (\ref{eq:distribution}),
(\ref{eq:con1}), (\ref{eq:con2}) and (\ref{eq:con3}), we can get
\begin{equation}\label{eq:m+}
    \overline{m}_{+}=\frac{1}{\beta}=\frac{1+\sqrt{1-r}}{r}\frac{M_{0}}{N}
\end{equation}
and
\begin{equation}\label{eq:m-}
    \overline{m}_{-}=-\frac{1}{\gamma}=\frac{1-r+\sqrt{1-r}}{r}\frac{M_{0}}{N}.
\end{equation}

It is seen that both $\overline{m}_{+}$ and $\overline{m}_{-}$
decrease as the required reserve ratio increases, and the value of
$\overline{m}_{+}$ is always larger than that of
$\overline{m}_{-}$ at the same required reserve ratio. These
results are illustrated by the solid lines in Figure $7$. They are
in good agreement with simulation results denoted by dots.

\subsection{The Velocity of Money Versus the Required Reserve Ratio}\label{holding}

The formula of the velocity of money will be deduced here. It is
known that the velocity of money is equal to the intensity of the
Poisson process from Equation (\ref{velocity1}). The intensity of
the Poisson process can be measured by average times a unit of
money takes part in trades in each round. This suggests that the
velocity of money is also the value of transaction volume in each
round $A$ divided by the money stock, i.e.,
\begin{equation}\label{velocity2}
    V=\frac{1}{T}=\frac{A}{M_0}.
\end{equation}
In order to obtain the expression of $V$ in terms of the required
reserve ratio, the analysis of $A$ is required at first.

For convenience in manipulation, traders are now classified into
two groups: the traders with positive monetary wealth and the ones
with non-positive monetary wealth, whose numbers are denoted by
$N_{+}$ and $N_{-}$ respectively. From the trading mode of our
model, it can be reckoned out that each trader participates in
trade averagely $2$ times in one round. In each transfer of money,
the probability of transferring a unit of money is $1/2$ for the
traders with positive monetary wealth, and it must be less than
$1/2$ for the traders with non-positive monetary wealth, for
borrowing may fail due to the limitation of required reserves. Let
$\omega$ denote this probability, from the detailed balance
condition which holds in steady state, we have
\begin{equation}\label{state}
   {\omega}p_{-}(m_{1})p_{+}(m_{2})=\frac{1}{2}p_{-}(m_{1}-1)p_{+}(m_{2}+1).
\end{equation}
Substituting the expressions of monetary wealth distribution
(\ref{eq:distribution}) into Equation (\ref{state}), we obtain
\begin{equation}\label{omega}
    \omega=\frac{1}{2}e^{-\frac{1}{\overline{m}_{+}}-\frac{1}{\overline{m}_{-}}}.
\end{equation}
Thus the total trade volume in each round on average can be
expressed as
\begin{equation}\label{trade}
    A=N_{+}+N_{-}e^{-\frac{1}{\overline{m}_{+}}-\frac{1}{\overline{m}_{-}}}.
\end{equation}
Substituting Equation (\ref{trade}) into (\ref{velocity2}), the
velocity of money can be given by
\begin{equation}\label{velocity3}
    V=\frac{N_{+}}{M_{0}}+\frac{N_{-}}{M_{0}}e^{-\frac{1}{\overline{m}_{+}}-\frac{1}{\overline{m}_{-}}}.
\end{equation}
Since in steady state the number of traders whose monetary wealth
changes from $0$ to $1$ is equal to that of traders whose monetary
wealth changes from $1$ to $0$, we have the following approximate
relation
\begin{equation}\label{state1}
   {\omega}N_{0}{\frac{N_{-}}{N}}+{\frac{1}{2}}N_{0}{\frac{N_{+}}{N}}={\frac{1}{2}}N_{1}{\frac{N_{+}}{N}}+{\frac{1}{2}}N_{1}{\frac{N_{-}}{N}},
\end{equation}
where $N_{1}=N_{0}e^{-\frac{1}{\overline{m}_{+}}}$ is the number
of traders with monetary wealth $1$. The left-hand side of
Equation (\ref{state1}) represents the number of traders whose
monetary wealth changes from $0$ to $1$ and the right-hand side
denotes the number of traders whose monetary wealth changes from
$1$ to $0$. Substituting Equation (\ref{omega}) into
(\ref{state1}) and taking $N=N_{+}+N_{-}$ into account yield
\begin{equation}\label{N_+}
    N_{+}={\frac{e^{\frac{1}{\overline{m}_{-}}}-1}{e^{\frac{1}{\overline{m}_{+}}+\frac{1}{\overline{m}_{-}}}-1}}N
\end{equation}
and
\begin{equation}\label{N_-}
    N_{-}={\frac{e^{\frac{1}{\overline{m}_{+}}+\frac{1}{\overline{m}_{-}}}-e^{\frac{1}{\overline{m}_{-}}}}{e^{\frac{1}{\overline{m}_{+}}+\frac{1}{\overline{m}_{-}}}-1}}N.
\end{equation}
Combining Equations (\ref{velocity3}), (\ref{N_+}) and
(\ref{N_-}), we can obtain
\begin{equation}\label{velocity_final}
   V= {\frac{N}{M_{0}}}e^{-\frac{1}{\overline{m}_{+}}}.
\end{equation}
Figure $8$ shows the relationships between the velocity of money
and the required reserve ratio, from simulation results and from
Equation (\ref{velocity_final}) respectively. By measuring latency
time for different required reserve ratios, the corresponding
velocities of money are obtained from Equation (\ref{velocity1}).
From Figure $8$, it is seen that the velocity of money has an
inverse relation with the required reserve ratio. This can be
interpreted in this way. In each round, if every pair of traders
could fulfill their transfer of money, the trade volume would be
$N$ in our model. However, in each round some transfers are
cancelled because the payers with non-positive monetary wealth may
not get loans from the bank. As indicated by Equation
(\ref{velocity_final}), the average realized transfer ratio can be
expressed in the form of $e^{-\frac{1}{\overline{m}_{+}}}$, which
decreases as the required reserve ratio increases. Thus the trade
volume in each round decreases, and as a result the velocity of
money decreases.

\section{Conclusion}
In this paper, in order to see how money creation affects the
statistical mechanics of money circulation, we develop a random
transfer model of money by introducing a fractional reserve
banking system. In this model, the monetary aggregate is
determined by the monetary base and the required reserve ratio.
Computer simulations show that the steady monetary wealth
distribution follows asymmetric Laplace type and latency time of
money obeys exponential distribution regardless of the required
reserve ratio. The distribution function of monetary wealth in
terms of the required reserve ratio is deduced. Likewise, the
expression of the velocity of money is also presented. These
theoretical calculations are in quantitative agreement with the
corresponding simulation results. We believe that this study is
helpful for understanding the process of money creation and its
impacts in reality.

\section*{Acknowledgments}
This research was supported by the National Science Foundation of
China under Grant No. 70371072 and 70371073. The authors are
grateful to Thomas Lux for comments, discussions and helpful
criticisms.

\newpage

\section*{Figure Captions}
\begin{description}
    \item[Figure 1] Time evolution of the monetary aggregate for the
required reserve ratio $r=0.8$. The vertical line denotes the
moment at which the monetary aggregate reaches a steady value.
    \item[Figure 2] Steady value of the monetary aggregate versus the
required reserve ratio obtained from simulation results (dots) and
from the corresponding analytical formula $M=M_0/r$ derived from
Equations (\ref{MA}) and (\ref{multi}) (continuous curve).
    \item[Figure 3] The moment at which the monetary aggregate reaches
a steady value versus the required reserve ratio.
    \item[Figure 4] The stationary distribution of monetary wealth for
the required reserve ratio $r=0.8$. It can be seen that the
distribution follows asymmetric Laplace distribution from the
inset.
    \item[Figure 5] The stationary distribution of latency time for
the required reserve ratio $r=0.8$. The fitting in the inset
indicates that the distribution follows an exponential law.
    \item[Figure 6] The stationary distributions of monetary wealth
for different required reserve ratios. Note that the probability
has been scaled by the corresponding maximum value.
    \item[Figure 7] $\overline{m}_{+}$ (upper) and $\overline{m}_{-}$
(lower) versus the required reserve ratio obtained from simulation
results (dots) and from the corresponding analytical formulas
(continuous curves) given by Equations (\ref{eq:m+}) and
(\ref{eq:m-}) respectively.
    \item[Figure 8] The velocity of money versus the required reserve
ratio obtained from simulation results (dots) and from the
corresponding analytical formula (continuous curve) given by
Equation (\ref{velocity_final}).
\end{description}

\begin{figure}
\includegraphics[width=\textwidth]{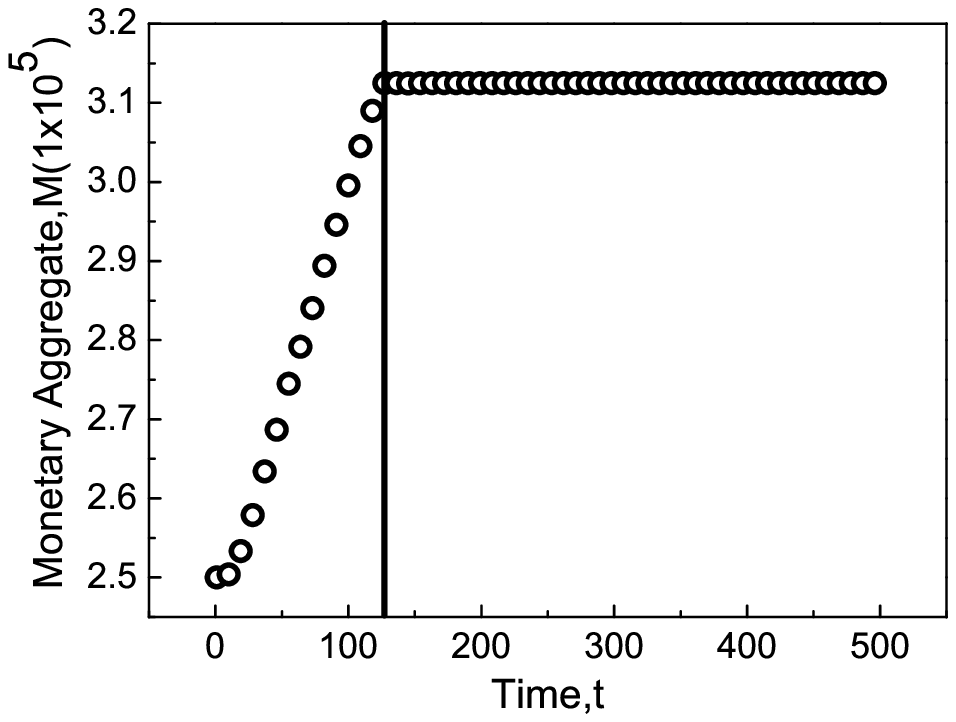}
\centering{\Large{Fig. 1}}
\end{figure}
\clearpage
\begin{figure}
\includegraphics[width=\textwidth]{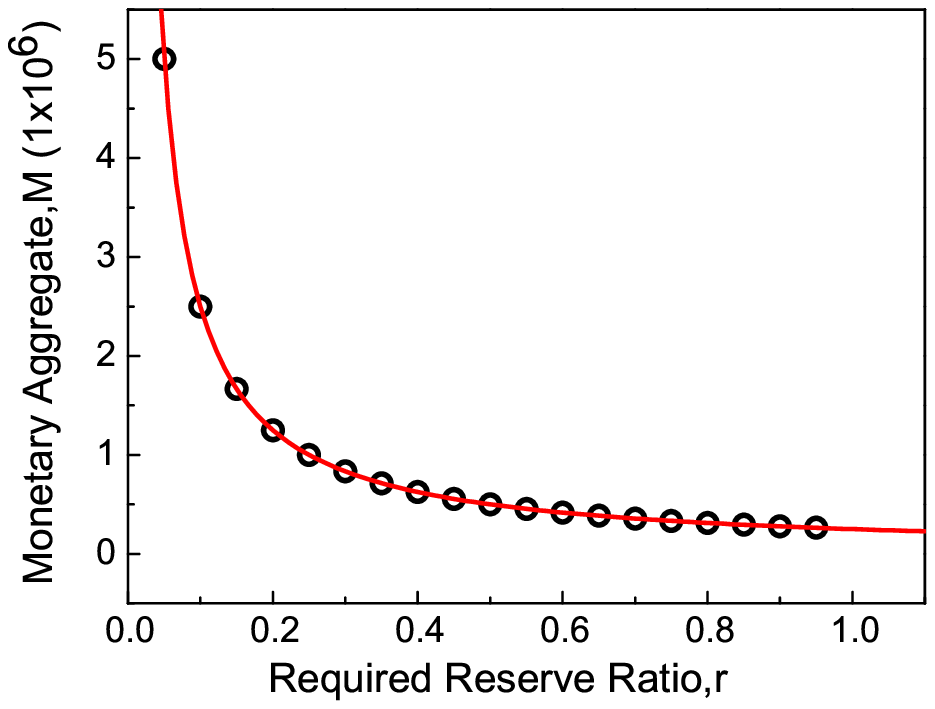}
\centering{\Large{Fig. 2}}
\end{figure}
\clearpage
\begin{figure}
\includegraphics[width=\textwidth]{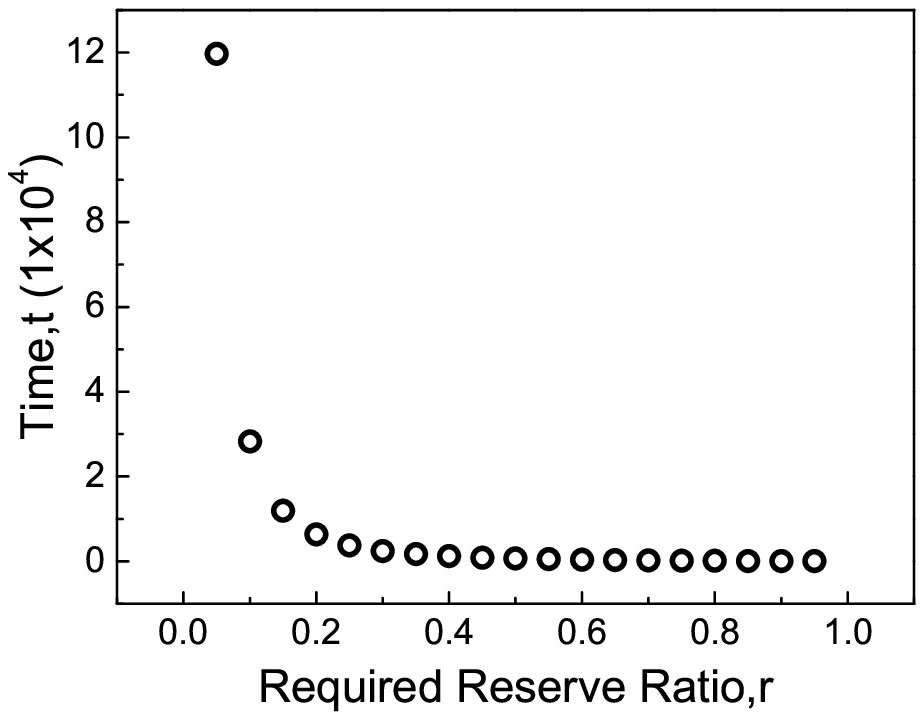}
\centering{\Large{Fig. 3}}
\end{figure}
\clearpage
\begin{figure}
\includegraphics[width=\textwidth]{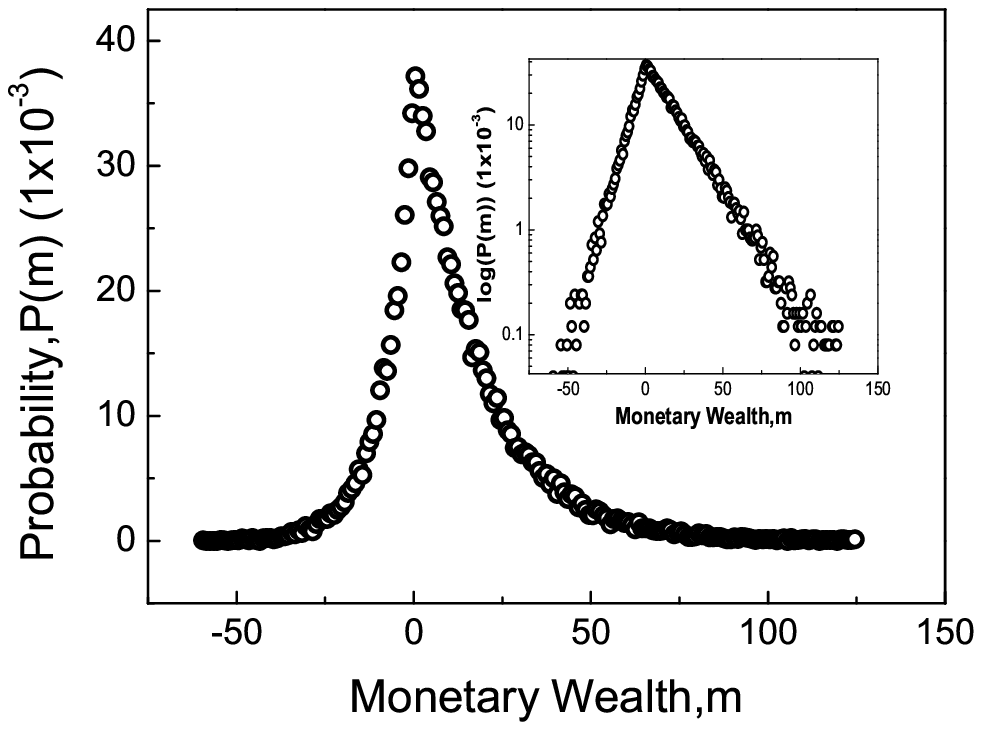}
\centering{\Large{Fig. 4}}
\end{figure}
\clearpage
\begin{figure}
\includegraphics[width=\textwidth]{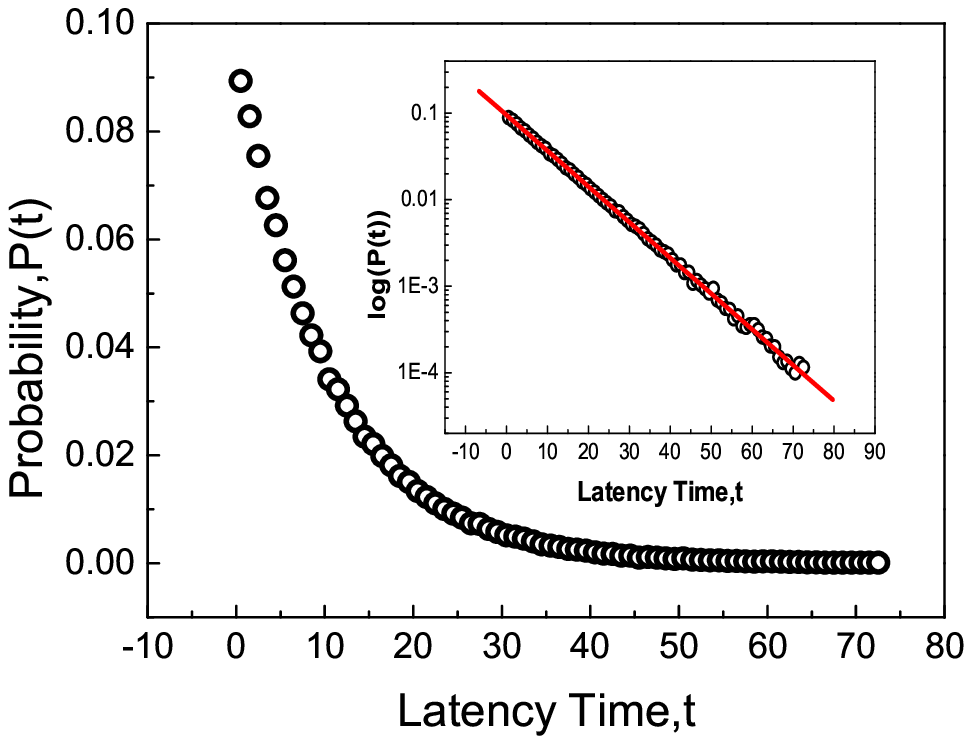}
\centering{\Large{Fig. 5}}
\end{figure}
\clearpage
\begin{figure}
\includegraphics[width=\textwidth]{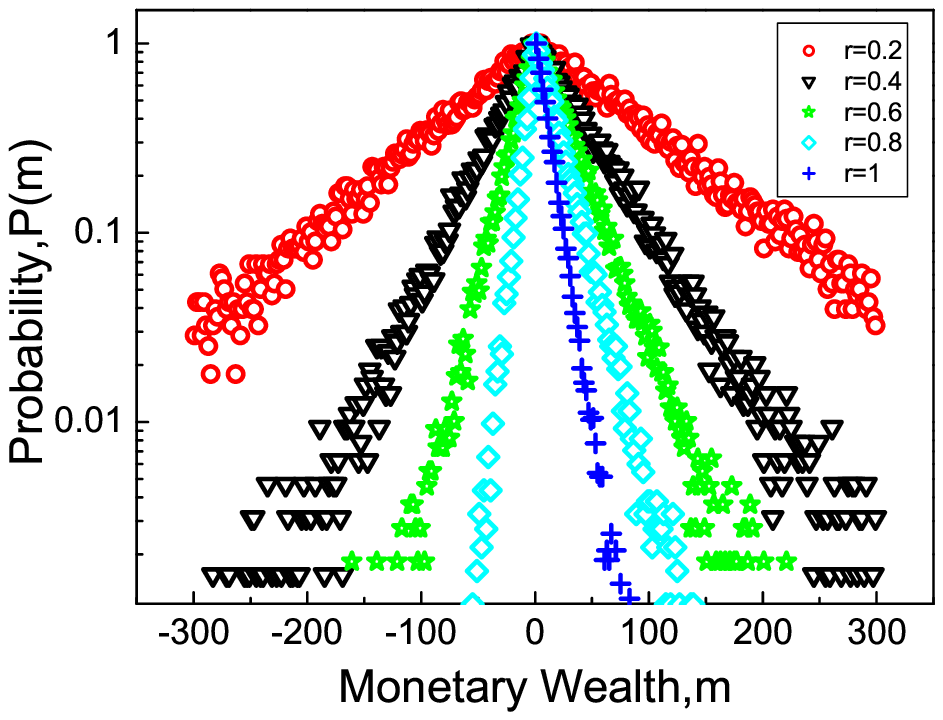}
\centering{\Large{Fig. 6}}
\end{figure}
\begin{figure}
\includegraphics[width=\textwidth]{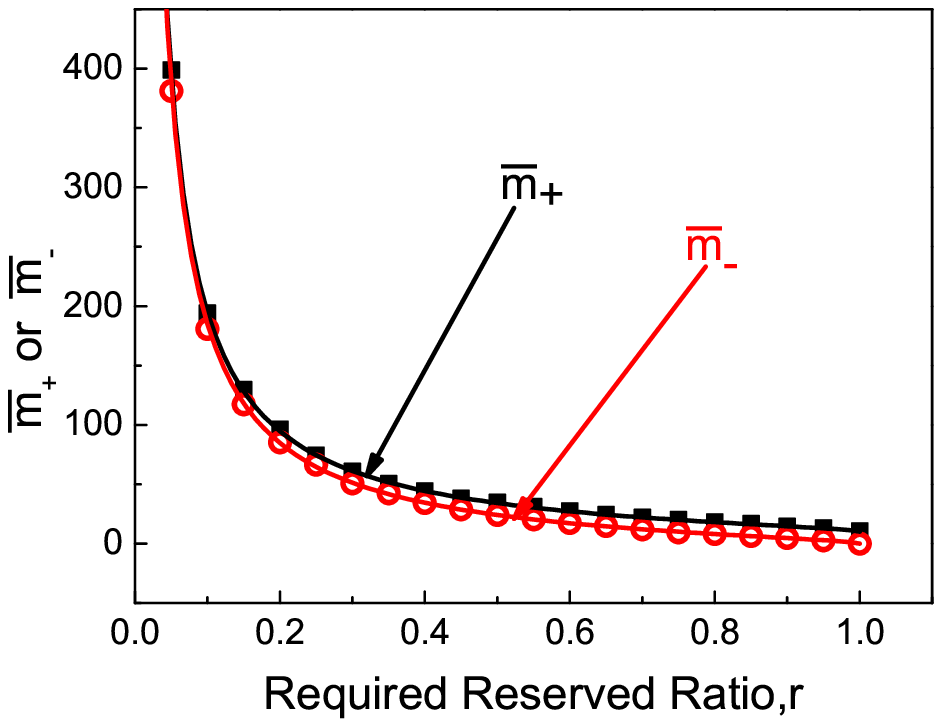}
\centering{\Large{Fig. 7}}
\end{figure}
\begin{figure}
\includegraphics[width=\textwidth]{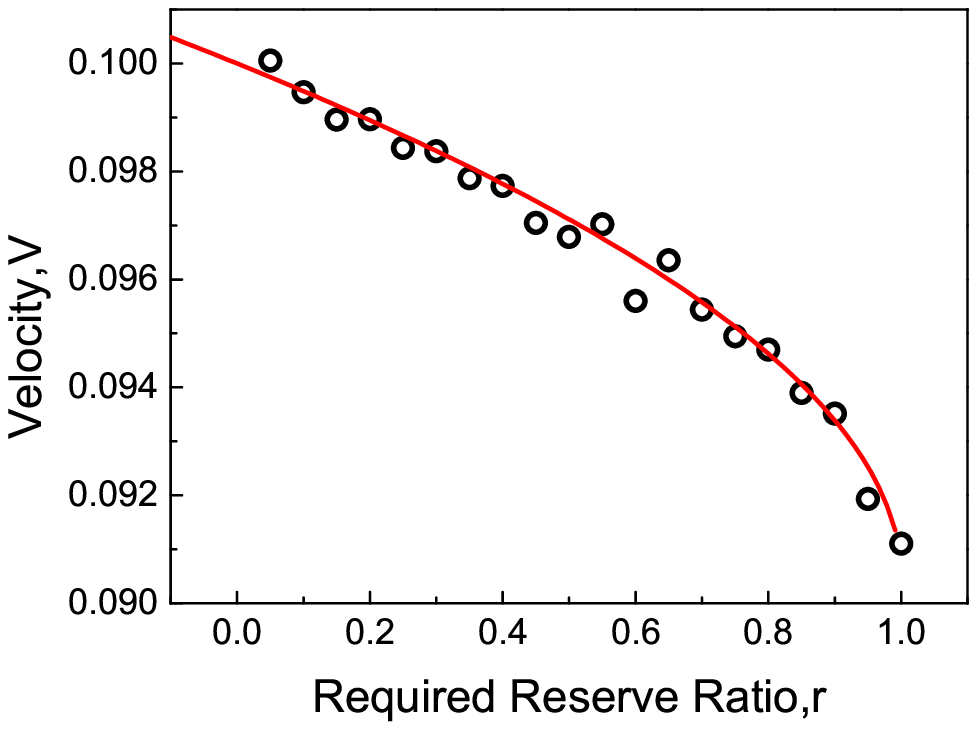}
\centering{\Large{Fig. 8}}
\end{figure}

\end{document}